\documentclass[prd,aps,twocolumn,superscriptaddress,nofootinbib,amsmath,amssymb,aps]{revtex4-1}

\usepackage{bbold}
\usepackage{float}
\usepackage{hhline}
\usepackage[table]{xcolor}
\usepackage{array,multirow}

\usepackage{graphicx}% Include figure files
\usepackage{bm}% bold math
\usepackage{combelow}
\usepackage{braket}
\usepackage{slashed}

\newcommand{\beqn}{\begin{eqnarray}}
\newcommand{\eeqn}{\end{eqnarray}}
\newcommand{\beqs}{\begin{subequations}}
\newcommand{\eeqs}{\end{subequations}\\[-3.5mm]\noindent}
\newcommand{\eq}[1]{(\ref{#1})}
\newcommand{\bs}{\boldsymbol}

\newcommand{\avr}[1]{{\left\langle #1 \right\rangle}}

\newcommand{\GeV}{\,{\mathrm{GeV}}}
\newcommand{\MeV}{\,{\mathrm{MeV}}}

\begin{document}

\title{Hyperon--anti-hyperon polarization asymmetry in relativistic heavy-ion collisions \newline as an interplay between chiral and helical vortical effects}

\author{Victor E. Ambru\cb{s}}
\affiliation{Department of Physics, West University of Timi\cb{s}oara, Bd.~Vasile P\^arvan 4, Timi\cb{s}oara 300223, Romania}
\affiliation{Institut f\"ur Theoretische Physik, Johann Wolfgang Goethe-Universit\"at, Max-von-Laue-Strasse 1, D-60438 Frankfurt am Main, Germany}
\author{M. N. Chernodub}
\affiliation{Institut Denis Poisson UMR 7013, Universit\'e de Tours, Tours, 37200, France}
\affiliation{Pacific Quantum Center, Far Eastern Federal University, 690950 Vladivostok, Russia}

\date{\today}

\begin{abstract}
We argue that the enhancement in the spin polarization of anti-hyperons compared to the polarization of the hyperons in noncentral relativistic heavy-ion collisions arises as a result of an interplay between the chiral and helical vortical effects. The chiral vortical effect generates the axial current of quarks along the vorticity axis while the recently found helical vortical effect generates the helicity flow -- the projection of the quark's polarization vector onto its momentum -- along the same axis. 
For massless fermions, the helical charge corresponds to a difference in the contributions of particles and anti-particles to the axial charge. Assuming that the spin of light quarks transfers to the strange quarks via the vector kaon states (``the spin-vector dominance''), we are able to describe the ratio of the (anti)hyperon spin polarizations, obtained by the STAR group, without fitting parameters. We also argue that the helical vortical effect dominates over the chiral vortical effect and the chiral magnetic effect in the generation of the electric current.
\end{abstract}

\maketitle

\section{Introduction}

Over the past decades, the experiments of relativistic heavy ion collisions at RHIC and LHC have served as an excellent arena to study the properties of quantum matter in extreme conditions. In particular, it was shown that, for sufficiently large collision energies, the medium created behind the colliding nuclei evolves from an initial glasma state into the quark-gluon plasma (QGP)~\cite{Heinz:2001xi,Gyulassy:2004zy,Shuryak:2004,Jacak:2012dx}.
This state of matter persists until the local temperature drops below a critical value, under which the quark and gluon degrees of freedom are reconfined within hadrons. Quite remarkably, the QGP appears to behave like a nearly-perfect fluid~\cite{Romatschke:2007mq,Heinz:2013th}, exhibiting the lowest known value for the ratio between shear viscosity and entropy density (an estimation based on the AdS/CFT correspondence is $\eta / s = 1 / 4\pi$~\cite{Son:2006em}). 
Its constituents achieve early-time thermal equilibrium, as supported by evidence from elliptic flow measurements~\cite{Ackermann:2000tr} (perhaps due to the decay of non-hydrodynamic modes to an attractor solution~\cite{Heller:2015dha}), such that a macroscopic hydrodynamic description of the QGP phase is appropriate~\cite{Florkowski:2017olj}. 

The spin-orbit coupling, inherent in the Dirac equation, leads to a polarization of the fermion spins with respect to the direction of the collective angular momentum. The transport consequences of this effect were highlighted more than four decades ago by Vilenkin, who pointed out that rotating (Kerr) black holes generate a net neutrino current directed along the axis parallel to the black hole's angular momentum~\cite{Vilenkin}. This phenomenon constitutes one of the chiral vortical effects  which were later understood in the framework of anomalous hydrodynamics of relativistic vortical fluids~\cite{Son:2009tf}.

It is commonly anticipated that in noncentral heavy-ion collisions, the produced quark-gluon plasma should polarize the spins of quarks and anti-quarks in the direction of the global angular momentum of the colliding ions. Perturbative calculations~\cite{Liang:2004ph} indicate that quarks and anti-quarks tend to align their spins in the direction of the local vorticity of a rotating fluid. The transfer of polarization proceeds via a scattering process with a parton in the rotating hot nuclear fluid.

The evidence of vorticity in QGP was reported by the STAR collaboration~\cite{STAR:2017ckg,Adam:2018ivw} based on the data collected during the experiments performed at the RHIC facility in Brookhaven National Laboratory in USA. More specifically, the experimental results indicate that the quark-gluon medium, formed in non-central collisions, exhibits a global vorticity, which is measured based on a specific decay of spin-polarized $\Lambda$-hyperons and $\bar\Lambda$ anti-hyperons. 

The analysis of the hyperon decays provides us with useful information on the spin properties of quark-gluon plasma because of the relatively high production cross-section of these baryons~\cite{Blume:2011}, as well as the ``self-analyzing'' property~\cite{ref:book} of their spin polarization due to the characteristic decay mode $\Lambda \to p \pi^-$ with a large branching ratio of 64\%. The self-analysis uncovers the spin polarization by preferentially emitting a daughter proton in the spin direction of $\Lambda$ and a daughter anti-proton opposite to the spin direction of $\bar\Lambda$. If $\theta^*$ is the angle between the momentum $\bm{p}_p^*$ and the hyperon polarization vector $\bm{\mathcal{P}}_H$ in the rest frame of $H = \Lambda, {\bar \Lambda}$, then
\begin{equation}
 \frac{d N_{H}}{d\cos\theta^*} = \frac{1}{2}(1+ \alpha_H |\bm{\mathcal{P}}_H| 
 \cos\theta^*),
 \label{eq:STAR1}
\end{equation}
where $\alpha_\Lambda = -\alpha_{\bar \Lambda} = 0.642 \pm 0.013$ represent the decay parameters for $\Lambda$ and ${\bar \Lambda}$. When averaging the polarization over the total number of events associated with one collision, the total polarization vector is required by symmetry to be parallel to the direction of the system's global angular momentum vector, $\hat{\bm{J}}_{\mathrm{sys}}$. The value of this global polarization is computed using
\begin{equation}
 \overline{\mathcal{P}}_H = \braket{\bm{P}_H \cdot \bm{\hat{J}}_{\rm sys}},
 \label{eq:STAR2}
\end{equation}
where the bar on $\mathcal{P}_H$ denotes an average over events.

The experimental results reveal nonvanishing spin polarizations of $\Lambda$ and $\bar\Lambda$ hyperons thus indicating the presence of a highly vortical quark-gluon fluid that emerge in non-central heavy-ion collisions. As the collision energy lowers, the spin polarizations of both $\Lambda$ and $\bar\Lambda$ grow. The growth is not, however, identical for these hyperons: at lower energies, the polarization of $\bar\Lambda$ is substantially higher than the polarization of $\Lambda$~\cite{STAR:2017ckg}.

The puzzling effect of the splitting in the polarizations has attracted a significant attention of the community. 
The effect could indicate a possible role of the magnetic field~\cite{Becattini:2016gvu}, which is, however, generally expected to be vanishing at the freeze-out (see the recent overview~\cite{ref:Niida:2020}).  While the lifetime of magnetic field may be enhanced by a high electric conductivity of quark-gluon plasma~\cite{McLerran:2013hla} and rotational effect of charged fluid~\cite{Guo:2019mgh}, the observed splitting may have also other explanations including different freeze-out conditions for hyperons and anti-hyperons~\cite{Vitiuk:2019rfv},
and effects of the axial and mixed axial-gravitational anomalies~\cite{Baznat:2013zx,Sorin:2016smp,Baznat:2017jfj}.
At suitable choice of parameters, the splitting in the hyperons' and anti-hyperons' spins may also be explained in an effective theory of quark interactions mediated by massive vector and scalar bosons~\cite{Csernai:2018yok}

We argue that the observed splitting in polarizations can occur due to an interference between the chiral vortical effect and the new helical vortical effect~\cite{Ambrus:2019ayb,Ambrus:2019khr}. 

The structure of our paper is as follows. In Section~\ref{sec:transport} we briefly overview the chiral and helical transport effects in the quark's vortical fluid, and stress the difference between the chirality and helicity of quarks. In Section~\ref{sec:mechanism} we propose the mechanism of the (anti)hyperon spin polarization via chiral and helical vortical effects. In Section~\ref{sec:ratio} we show that the ratio of the spin polarizations of the anti-hyperons and hyperons is determined only by chiral and helical vortical conductivities evaluated as the chemical freeze-out. Even though the theoretical prediction for this ratio contains no adjustable phenomenological parameters, it agrees excellently with the experimental data of the STAR collaboration. In Section~\ref{sec:electric} we stress the importance of the helical degrees of freedom demonstrating that at large collision energies $\sqrt{s_{{NN}}} = 200 \GeV$, the helical vortical effects generate an electric current of much stronger magnitude as compared with both the chiral vortical effect and the chiral magnetic effect. The last Section is devoted to our conclusions.

\section{Transport in vortical fluid}
\label{sec:transport}

\subsection{Chiral and helical vortical effects}

Vilenkin's findings represent the first evidence of the chiral vortical effect, which predicts that the quark fluid at finite temperature $T$ generates a net chiral charge current along the vorticity~$\bm{\omega}$:
\begin{equation}
 \bm{J}_A = \sigma_A \bm{\omega}, \qquad 
 \sigma_A = \frac{T^2}{6} + \frac{\mu_B^2}{18\pi^2}.
 \label{eq:sigmaA}
\end{equation}
Here we adapted the notations to incorporate the relation $\mu_V = \frac{1}{3} \mu_B$ between the vector chemical potential for quarks $\mu_V$ and the baryonic charge, $\mu_B$ (for a review, see~\cite{Kharzeev:2015znc}).

The coefficient in front of the first term in the conductivity~\eq{eq:sigmaA} originates from the mixed axial-gravitational anomaly. It has been probed in the first-principle lattice simulations of the so-called axial magnetic effect, which has the same conductivity as the chiral vortical effect~\cite{Braguta:2013loa,Braguta:2014gea}. The coefficient turned out substantially smaller than its predicted value of 1/6, which may, however, be a result of the fermion quenching used in the numerical simulations and needs a further check. The second term in the conductivity~\eq{eq:sigmaA} is not renormalized by quantum corrections as it comes from the axial anomaly which is exact in one loop. Below, we will use the axial conductivity as given in Eq.~\eq{eq:sigmaA}.

Besides chirality, the polarization of free fermions can be characterized by the helicity of the fermionic current~\cite{Ambrus:2019khr,Ambrus:2019ayb}. Similarly to its chiral counterpart, the helical vortical effect is a mechanism through which a helicity charge current is generated along the vorticity axis:
\begin{equation}
 \bm{J}_H = \sigma_H \bm{\omega}, \qquad 
 \sigma_H = \frac{2 \ln 2}{3\pi^2} \mu_B T + O\left(\frac{\mu_B^3}{T}\right).
 \label{eq:sigmaH}
\end{equation}
The above helicity current is often overlooked in estimations of polarization. In Eq.~\eq{eq:sigmaH}, the helical vortical conductivity $\sigma_H$ is written in the high-temperature limit and the sub-leading terms of the series are not presented. Our calculations show that these terms are negligible in the regions of studied collision energies. In what follows, we also ignore a small isospin charge of the quark-gluon plasma which appears due to a light imbalance in densities of $u$- and $d$-quarks in the plasma.

\subsection{Helicity vs. chirality}

The helicity is often misinterpreted as the chirality. However, the very definitions of the corresponding four-currents are different:
\begin{align}
 J^\mu_A = \overline{\psi} \gamma^\mu \gamma^5 \psi, 
\qquad
 J^\mu_H = \overline{\psi} \gamma^\mu h \psi + \overline{h \psi} \gamma^\mu \psi.
 \label{eq:J}
\end{align}
The helicity operator 
\beqn
h = \frac{{\bs s}\cdot {\bs p}}{p} \equiv \frac{1}{2}\gamma^5 \mathrm{sign}\,({\hat H}),
\label{eq:h}
\eeqn
is the projection of the spin operator $s^i = \frac{1}{2} \varepsilon^{0ijk} \Sigma_{jk}$
(a spatial part of the covariant spin tensor $\Sigma^{\mu\nu} = \frac{i}{4}[\gamma^\mu,\gamma^\nu]$) into the direction of momentum. Notice that in the definition of the helical current in Eq.~\eq{eq:J}, the helicity operator~\eq{eq:h} enters twice, so that the spin's 1/2 factor disappears.

For a massless fermion, employed in the definition~\eq{eq:h}, the helicity can be expressed via the sign of the particle's Hamiltonian $\hat H$~\cite{Ambrus:2019ayb}. The latter property highlights the interrelation between the helicity and the chirality: the helicity is the difference between the axial currents carried by particles and anti-particles.

In this paper, we aim to show that taking into account the two transport laws of chiral~\eqref{eq:sigmaA} and helical~\eqref{eq:sigmaH} degrees of freedom, it is possible to reproduce the experimental data for the ratio of the hyperon polarizations, as measured in heavy-ion collisions.

\section{Hyperon spin polarization via chiral and helical vortical effects}
\label{sec:mechanism}

\subsection{Basic idea of the mechanism}

We will consider the following simplified mechanism of the spin polarizations. The vortical fluid of QGP created in a non-central heavy-ion collision polarizes the spins of light quarks in early stages of the collision. The light quarks pass then their spin polarization to the strange quarks via intermediate vector kaon resonances (we call this mechanism ``the spin-vector dominance''). The polarization of the strange quarks is then observed in experiment as the spin polarizations of hyperons. The polarization of the light quarks via the chiral and helical vortical effects is enough to explain the ratio in the spin polarizations of the hyperons without fitting parameters. 

In order to make our derivation algebraic and as simple as possible, we focus on the dominant effects of the mechanism. We ignore the {\it direct} polarization of the massive strange quarks in the vortical fluid which proceeds without scattering on already polarized light quarks. The strangeness neutrality of the quark-gluon plasma dictates that the direct polarization mechanism would polarize the spins of $s$ and $\bar s$ in the same proportion, which would lead to identical polarizations of $\Lambda$ and $\bar \Lambda$. The latter suggestion is in contradiction with the experiments at low collision energies. Therefore, in our calculations, we exclude the direct polarization of the strange quarks which is expected to be small anyway~\cite{footnote:Kapusta}. 

The hypothesis, that the generated magnetic field is responsible for the observed splitting in polarization between the hyperons and anti-hyperons has been also studied~\cite{Becattini:2016gvu,Guo:2019joy} and it was estimated that the required magnetic field life-time ``not impossible''~\cite{Guo:2019joy}, while the current experimental results give the magnetic field at the chemical freeze-out consistent with zero~\cite{ref:Niida:2020}. In our study, we neglect the presence of the magnetic field and argue that the splitting can occur due to the helical vortical effects.

\subsection{Vector and axial charges, and quark's helicity}

We study the polarization properties of the hyperons with respect to the direction of the global polarization vector $\bm{\hat{J}}_{\rm sys}$ of a noncentral heavy-ion collision. We start our analysis with the spin polarization of the light, $u$ and $d$, relativistic quarks. Since the freeze-out temperature is much higher than the current masses of these quarks, we treat the light quarks as massless fermions.

A single quark may be characterized according to its vector charge content (carried by particles $j$ and anti-particles ${\bar j}$), its chirality (distinguished between right-chiral $R$ and left-chiral $L$) as well as its helicity (right-helical $\uparrow$ and left-helical $\downarrow$). These three quantities are constrained for a single particle or anti-particle because the chirality of a quark (an anti-quark) is the same as (opposite to) its helicity. The exact correspondence is lost for an ensemble of particles where the chirality and helicity become independent characteristics of the particle ensemble~\cite{Ambrus:2019khr}.

\begin{figure}[t!]
\begin{center}
\includegraphics[scale=0.475]{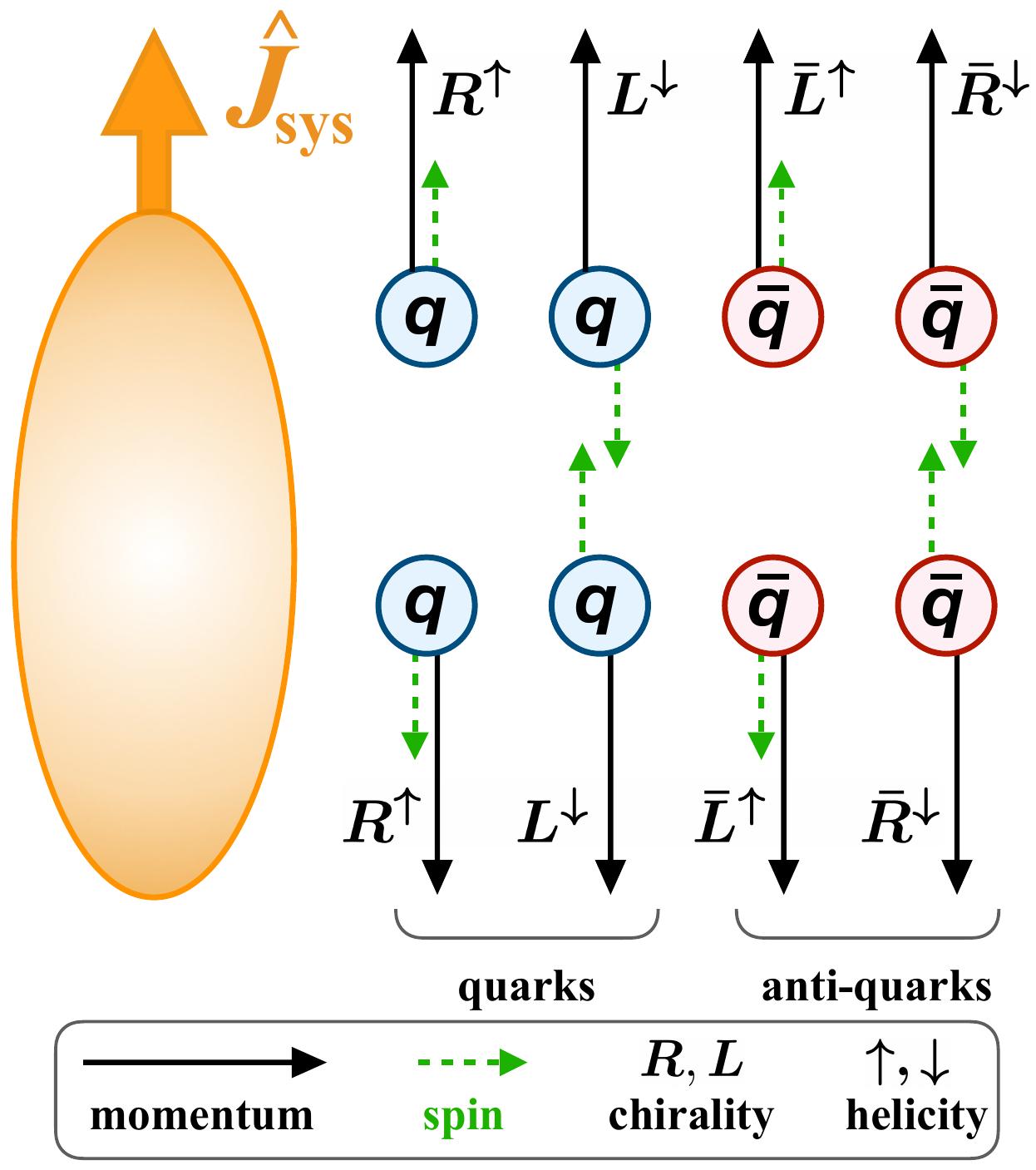}
\end{center}
\vskip -3mm
\caption{Schematic picture of spin polarization, momentum, chirality and helicity of light quarks and anti-quarks projected onto the direction of the global angular momentum $\bm{\hat{J}}_{\rm sys}$ of a vortical quark-gluon plasma.}
\label{fig:notations}
\end{figure}

In Fig.~\ref{fig:notations} we illustrate the interrelations between spin polarization, momentum, chirality, and helicity of light quarks and anti-quarks. The notations are self-explaining: for example, ${\bar J}_\downarrow^R$ denotes the current of right-chiral ($R$) anti-quarks (${\bar j}$) which carry, necessarily, the left-handed helicity ($\downarrow$).

\subsection{Axial and helical currents, and spin polarization}

For notation simplicity, we omit below the vector indices of the currents, and consider only the projection of the currents to the axis of rotations, 
\beqn
J_\ell = {\bs J}_\ell \cdot {\bs n}, 
\qquad
{\bs n} = \frac{\bm{\hat{J}}_{\rm sys}}{{\hat{J}}_{\rm sys}} \equiv \frac{{\bs \omega}}{\omega}, 
\qquad \ell = V,A,H,
\label{eq:current:projection}
\eeqn
where the unit vector ${\bs n}$ points along the global vorticity directed out of the reaction plane and the index $\ell$ in Eq.~\eq{eq:current:projection} denotes the type of the charge (vector, axial, and helical, respectively) carried by the current. The vorticity ${\bs \omega}$ of rotating quark fluid is directed along its global angular momentum ${\bm{\hat{J}}_{\rm sys}}\| {\bs \omega}$.

We do not indicate the flavor index $f=u,d$ of the quark's current because they contribute on an equal basis. We also do not show the brackets of the mean expectation values, so that $J_\ell = \avr{J_\ell} \equiv \avr{{\bs J}_\ell \cdot {\bs n}}$ etc. 

We have four linearly-independent quantities which characterize the charge/particle content of the (projected) fermionic current:
\begin{itemize}

\item The total current which accounts for the flow of all particles and anti-particles:
\beqn
J_{\mathrm{tot}} = J_\uparrow + J_{\downarrow} + {\bar J}_\uparrow + {\bar J}_{\downarrow}.
\label{eq:N:tot}
\eeqn

\item The vector (electric) current given by the total number of particles minus the total number of anti-particles regardless of their helicities:
\beqn
J_V = J_\uparrow + J_{\downarrow} - {\bar J}_\uparrow - {\bar J}_{\downarrow}. 
\label{eq:N:V}
\eeqn

\item The axial (chiral) current which equals to the flow of the total number of particles and anti-particles with right-handed helicity minus the flow of the total number of particles and anti-particles with left-handed helicity:
\beqn
J_A =J_\uparrow + {\bar J}_\uparrow - J_{\downarrow} - {\bar J}_{\downarrow}.
\label{eq:N:A}
\eeqn

\item The helical current (the helicity flow) which is given by the linearly complementary quantity to all mentioned currents~\eq{eq:N:tot}, \eq{eq:N:V}, and \eq{eq:N:A}:
\beqn
J_H =J_\uparrow + {\bar J}_{\downarrow} - J_{\downarrow} - {\bar J}_\uparrow.
\label{eq:N:H}
\eeqn
The helical current has the meaning of the difference between the contributions to the axial current coming from particles and anti-particles~\cite{Ambrus:2019khr}.
\end{itemize}
Relations similar to Eqs.~\eq{eq:N:tot}--\eq{eq:N:H} are obviously also valid for the vector, axial and helical charge densities. 

% The relations, similar to Eqs.~\eq{eq:N:tot}--\eq{eq:N:H}, are obviously also valid for the vector, axial and helical charge densities. 

The total current~\eq{eq:N:tot}, which characterizes the quark's multiplicity flow, and the vector current~\eq{eq:N:V}, which gives the charge flow, are not relevant quantities for our aims since both these currents are neutral in their spin polarization contents. What is interesting for us is the sum and the difference between the axial and helical currents, which select the polarization of light quarks and anti-quarks, respectively:
\beqn
J_A + J_H = 2 (J_\uparrow - J_\downarrow), 
\qquad
J_A - J_H = 2 ({\bar J}_\uparrow - {\bar J}_{\downarrow}). \quad
\label{eq:j:sigma}
\eeqn
These relations indicate that the knowledge of the axial current $J_A$ is not enough to determine the spin polarizations of massless quarks.

\subsection{Currents and spin polarizations of light quarks}

The linear combinations of light-quark currents~\eq{eq:j:sigma} determine the spin polarizations of the light quarks and light anti-quarks, respectively. For example, the current $J_\uparrow$ carries the up-polarized spin to the upper hemisphere (if $J_\uparrow >0$) or to the lower hemisphere (if $J_\uparrow < 0$), depending on the sign of the current. 

The spins polarizations of light quarks $q = u,d$ and light anti-quarks ${\bar q} = {\bar u}, {\bar d}$ are related to the linear combinations of the currents~\eq{eq:j:sigma}:
\beqn
{\mathcal P}_{q} = \kappa_{q j} (J_\uparrow - J_\downarrow),
\ \
{\mathcal P}_{{\bar q}}
= \kappa_{{\bar q} {\bar j}} ({\bar J}_\uparrow - {\bar J}_{\downarrow}),
\quad
\kappa_{q j} = \kappa_{{\bar q} {\bar j}},
\qquad
\label{eq:q:j}
\eeqn
where $\kappa_{q j}$ and $\kappa_{{\bar q}{\bar j}}$ are positively-defined kinematic factors which do not distinguish the baryonic charge. As we neglect the masses of the light quarks $q = u,d$ and the presence of the magnetic field, there is no difference in spin polarization of light quarks of different flavors.

It is important to stress the crucial role of the helicity current for the correct evaluation of the spin polarization. First of all, let us make sure that the signs in the relations of Eq.~\eq{eq:q:j} take appropriately into account the spins of the particles that are emitted in upper and lower hemispheres. A quark with a positive helicity (i.e. with the spin oriented along the direction of motion) traveling to the upper (lower) hemisphere -- thus, with $J_\uparrow > 0$ ($J_\uparrow < 0$) and with $J_\downarrow = 0$ -- gives a positive (negative) contribution to ${\mathcal P}_{q}$ as it should be. At the same time, a quark with a negative helicity (i.e., with spin and momenta pointing in the opposite directions) moving to the upper (lower) hemisphere -- thus, with $J_\uparrow = 0$ and $J_\downarrow > 0$ ($J_\downarrow < 0$) -- gives a negative (positive) contribution to ${\mathcal P}_{q}$, as expected. The same calculation is true for the anti-quarks. 

Next, let us illustrate that the axial current alone cannot account for the spin polarization and that we need the helicity current. As an example, we take a quark and an anti-quark with a positive helicity (with the spin along the momentum) and the same magnitudes, but moving in opposite directions. We have $J_\uparrow = -{\bar J}_\uparrow \neq 0$ and $J_{\downarrow} = {\bar J}_{\downarrow} = 0$.
The positive-helicity quark (anti-quark), moving to the upper (lower) hemisphere gives a positive (negative) contribution to the total quark's (anti-quark's) spin polarization. As we just saw, our formulas~\eq{eq:q:j} correctly calculate the spins of quarks and anti-quarks due to the careful accounting for the helicity current. However, the axial current~\eq{eq:N:A} is identically zero for this simple configuration. Therefore, the axial current alone is insufficient for the separate computation of the spin polarizations of both particles and anti-particles.

%\vskip 1mm
%\paragraph{Spin-vector dominance: passing spin polarization from light quarks to strange quarks via vector kaon states.}
\subsection{Spin-vector dominance}

The rotating fluid polarizes the spin of the light $u$ and $d$ quarks. However the light-quark polarizations do not contribute to the total spin of the hyperons and anti-hyperons which are detected in the experiment. Moreover, according to the quark model, the contribution of the light $u$ and $d$ quarks to the total spin of the $\Lambda$ hyperon is strictly zero and the whole hyperon spin is carried by the $s$ quark only. First-principle numerical lattice calculations of the quark distribution functions of a $\Lambda$ hyperon confirm the qualitative validity of the quark model, giving $\Delta u_\Lambda = \Delta d_\Lambda = -0.02(4)$ for the contribution of the light quarks as compared to the longitudinal polarization $\Delta s_\Lambda = 0.68(4)$ of the $s$ quark~\cite{Gockeler:2002uh}. The same statement applies, of course, to the $\bar \Lambda$ antiparticle.

\begin{figure}[t!]
\begin{center}
\includegraphics[scale=0.275]{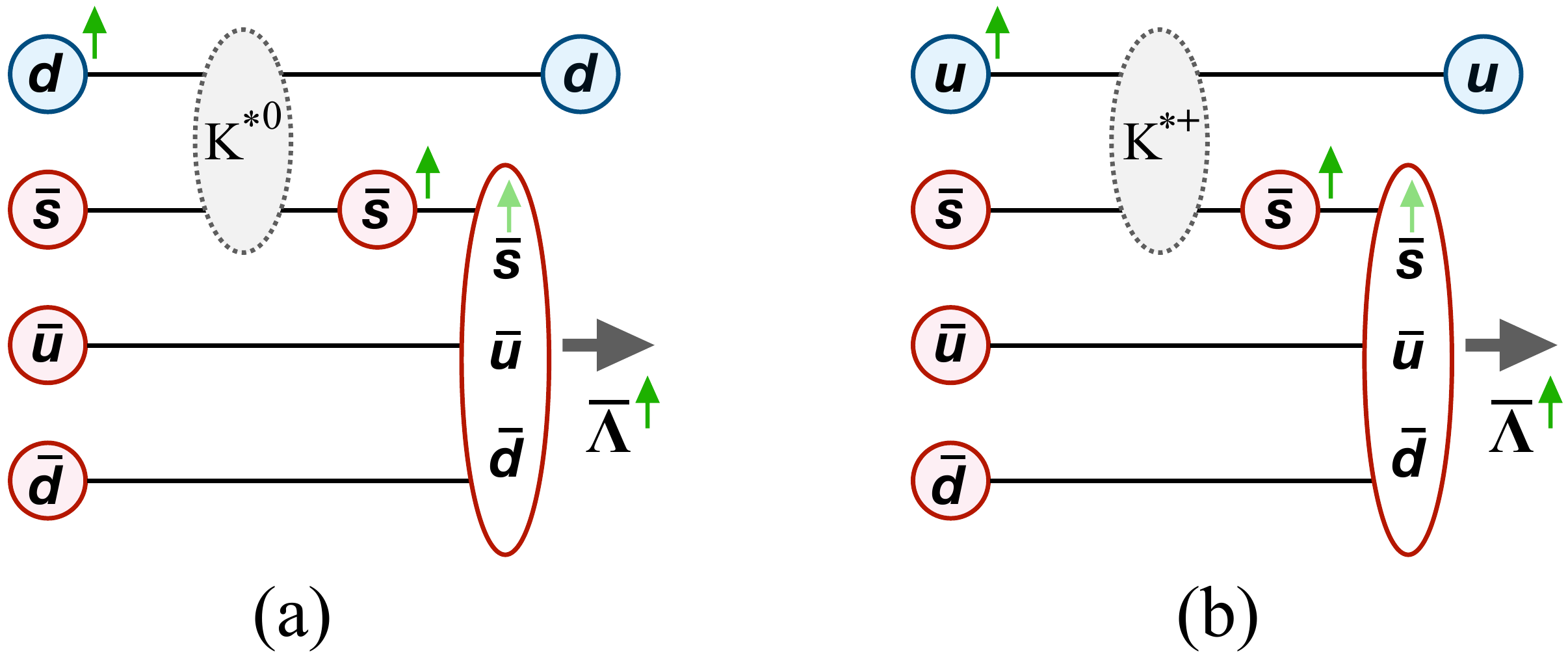}
\end{center}
\vskip -3mm
\caption{Polarization of a light quark is transferred, in an inelastic scattering via an intermediate vector kaon state -- for example, via (a) $K^{*0}= d{\bar s}$ and (b) $K^{*+}= u{\bar s}$ -- to an anti-strange quark and then to a final anti-hyperon~$\bar\Lambda$. Hyperons $\Lambda$ acquire their polarization from spin-polarized light anti-quarks in the $C$-conjugated processes (not shown).}
\label{fig:spin:transfer}
\end{figure}

Therefore, the polarized light quarks transfer their spin polarization to the strange quarks. As the perturbative transfer of polarization is inefficient~\cite{Kapusta:2019sad} we suggest that the spin transfer proceeds via spin-1 kaon vector states and a subsequent hadronization by the process of recombination to hyperons. In the vector $K^*$ kaons, the spins of the valence quark and anti-quark are aligned in the same direction. The polarized light (for example, $d$) quark scatters over an anti-strange quark $\bar s$ via the $K^{*0} = d {\bar s}$ resonant state. After the scattering, the $\bar s$ anti-quark picks the polarization of the $d$ quark. An illustration of these reactions is shown in Fig.~\ref{fig:spin:transfer}.

The polarizations of the strange quark and anti-quarks after the scattering process are, respectively, as follows:
\beqn
{\mathcal P}_{s}  = \kappa_{s{\bar q}} {\mathcal P}_{\bar q}, 
\qquad
{\mathcal P}_{\bar s}  = \kappa_{{\bar s} q} {\mathcal P}_{q}, 
\qquad\
\kappa_{s{\bar q}} = \kappa_{{\bar s} q},
\label{eq:s:q}
\eeqn
where $\kappa_{s{\bar q}}$ and $\kappa_{{\bar s} q}$ are the dynamical factors which characterize the efficiency of the spin-polarization transfers $q \to {\bar s}$ and ${\bar q} \to s$, correspondingly. These factors are equal to each other due to the charge-conjugation symmetry.

The spin polarizations of strange quarks and strange anti-quarks give us the spin polarizations of $\Lambda$ hyperons and $\bar\Lambda$ anti-hyperons, respectively:
\beqn
{\overline {\mathcal P}}_{\Lambda'} = \kappa_{\Lambda s} {\mathcal P}_{s}, 
\qquad
{\overline {\mathcal P}}_{{\bar\Lambda}'} = \kappa_{{\bar \Lambda} {\bar s}} {\mathcal P}_{\bar s}.
\qquad 
\kappa_{\Lambda s} = \kappa_{{\bar \Lambda} {\bar s}},
\label{eq:Lambda:s}
\eeqn
where we denoted ${\overline {\mathcal P}}_{\Lambda'} \equiv {\mathcal P}_{\Lambda}$ for notational similarity with Ref.~\cite{STAR:2017ckg}. The prime indicates that these polarizations are calculated for primary hyperons that are emitted directly from the quark-gluon fluid. We exclude the consideration of the feed-down effects~\cite{Becattini:2016gvu} where the primary polarizations may be diluted by (15\%-20\%) which falls within the experimental uncertainties~\cite{ref:Niida:2020}.

The dynamical factors $\kappa_{\Lambda s}$ and $\kappa_{{\bar \Lambda} {\bar s}}$ determine the production rate of the hyperons and any-hyperons, respectively. Since the strange quarks are already polarized, their hadronization in the hyperons will polarize the latter. The production factors are common to both particles and anti-particles due to the charge-conjugation symmetry. The spin polarization of $\Lambda$ corresponds to the parent $s$-quark's polarization in the same way as the spin polarization of $\bar\Lambda$ depends on the polarization of the $\bar s$ anti-quark. 

Bringing all equations~\eq{eq:sigmaA}, \eq{eq:sigmaH}, \eq{eq:current:projection}, \eq{eq:j:sigma}, \eq{eq:q:j}, \eq{eq:s:q}, and \eq{eq:Lambda:s} together, we get the spin polarizations of the hyperons due to axial and helical vortical effects:
\beqn
{\overline {\mathcal P}}_{\Lambda'} & = & \frac{1}{2} \kappa_{\Lambda s} \kappa_{s{\bar q}} \kappa_{{\bar q} {\bar j}} (\sigma_A - \sigma_H) \omega, 
\label{eq:int:1}\\
{\overline {\mathcal P}}_{{\bar \Lambda}'} & = & \frac{1}{2} \kappa_{{\bar \Lambda} {\bar s}} \kappa_{{\bar s} q} \kappa_{q j} (\sigma_A + \sigma_H) \omega,
\label{eq:int:2}
\eeqn
where the axial $\sigma_A$ and helical $\sigma_H$ vortical conductivities are given in Eqs.~\eq{eq:sigmaA} and \eq{eq:sigmaH}, respectively. The $\kappa$ factors for particles and anti-particles are pairwise equal to each other according to the $C$ symmetry mentioned above.

\section{Ratio of spin polarizations: theory vs experiment}
\label{sec:ratio}

The calculation of the factors $\kappa$ and $\omega$ in Eqs.~\eq{eq:int:1} and \eq{eq:int:2} may pose a significant challenge while being subjected, at the same time, to unwanted phenomenological and model-dependent uncertainties. In order to avoid these disadvantages, we consider the ratio of the polarizations,
\beqn
{\mathcal R}_{{\bar\Lambda}/\Lambda} = \frac{{\overline{\mathcal P}}_{{\bar\Lambda}'}}{{\overline{\mathcal P}}_{\Lambda'}},
\label{eq:R:definition}
\eeqn
for which the mentioned kinematic, dynamical and vortical factors disappear according to Eqs.~\eq{eq:int:1} and \eq{eq:int:2}. It is important to realize that the $\kappa$ factors contain also the relativistic contributions. While freeze-out can occur at different times for different space points, the local thermodynamic variables (temperature $T$ and baryon chemical potential $\mu_B$) should be the same. 

The vortical conductivities appearing in Eqs.~\eqref{eq:sigmaA} and \eqref{eq:sigmaH} have a weak dependence on the distance $\rho$ to the rotation axis measured in the reaction plane, since the local temperature ($T = T_0 \Gamma$) and the chemical potential ($\mu_B = \mu_0 \Gamma$) depend on the local Lorentz factor, $\Gamma = (1 - \rho^2 \Omega^2)^{-1/2}$. This relativistic factor becomes relevant only close to the speed of light surface (SLS), located at the distance $\Omega^{-1}$ from the rotation axis. For $\Omega \simeq 10^{22}\,{\rm s}^{-1}$, the SLS is located at $\sim 30\ {\rm fm}$, easily exceeding the diameter $d$ of the gold or lead nuclei ($d \sim 14\ {\rm fm}$). Thus, the dependence of the conductivities $\sigma_A$ and $\sigma_H$ on the distance to the rotation axis can be safely ignored. Furthermore, the local kinematic vorticity factor $\bm{\omega}$ appearing in Eqs.~\eqref{eq:sigmaA} and \eqref{eq:sigmaH} is the same for both particles and anti-particles. The operation of boosting from the proper frame to the laboratory frame is thus independent of the local values of the charge conductivities and can be factored out. Since the relativistic factors are identical for the quarks (hyperons) and anti-quarks (anti-hyperons) at the same expanding hypersurface, in the scope of our simplified approach, they disappear naturally in the ratio~\eq{eq:R:definition}.

The interplay between the chiral and helical vortical effects leads to the following prediction for the ${{\bar\Lambda}/\Lambda}$ polarization ratio~\eq{eq:R:definition}:
\beqn
{\mathcal R}_{{\bar\Lambda}/\Lambda} = \frac{\sigma_A + \sigma_H}{\sigma_A - \sigma_H}.
\label{eq:R:prediction}
\eeqn

As we already mentioned, we ignore the effect of a direct polarization of the strange quarks by the fluid vorticity. The vortical effects are not important for the strange quarks due to the strangeness neutrality of the quark-gluon plasma. The latter property implies that the strange chemical potential $\mu_s$ is globally zero in the high-temperature phase~\cite{Fu:2018qsk}. Therefore, the {\it direct} influence of the helical vortical effect on the polarization of the strange quarks is absent because the helical vortical conductivity for the strange quarks vanishes, $\sigma_H^{(s)} \sim \mu_s = 0$, according to Eq.~\eq{eq:sigmaH} with the substitution $\mu_B \to 3 \mu_s$.

\begin{figure}[t!]
\begin{center}
\includegraphics[scale=0.55]{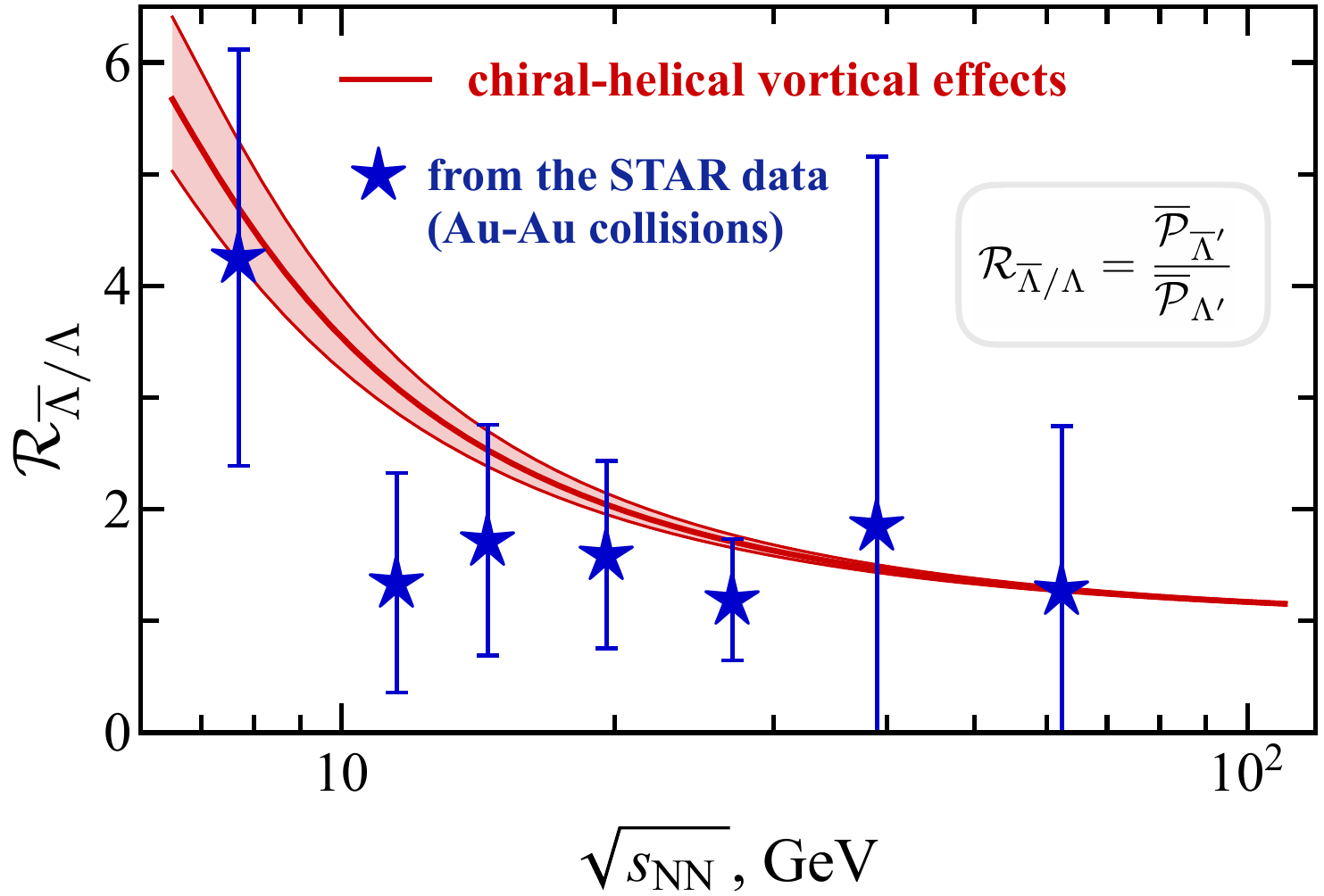}
\end{center}
\caption{Ratio ${\mathcal R}_{{\bar\Lambda}/\Lambda} {=}\, {\overline{\mathcal P}}_{{\bar\Lambda}'}/{\overline{\mathcal P}}_{\Lambda'}$ of the average global polarizations ${\overline{\mathcal P}}$ of anti-hyperons ${\bar\Lambda}$ and hyperons $\Lambda$, Eq.~\eq{eq:R:definition}, in noncentral Au-Au collisions. The experimental data of the STAR collaboration~\cite{STAR:2017ckg} is compared with the analytical prediction coming from the chiral-helical vortical effects~\eq{eq:R:prediction} with no fitting parameters used. The error bars of the data represent the statistical errors coming from the STAR experiment. The shadowed region corresponds to the confidence region of the analytical prediction coming from the chemical freeze-out parameters, Eq.~\eq{eq:parametrisation} and Table~\ref{tbl:freezeout}.}
\label{fig:R}
\end{figure}

The chiral~\eq{eq:sigmaA} and helical~\eq{eq:sigmaH} vortical conductivities entering the polarizations' ratio~\eq{eq:R:prediction} should be evaluated at the chemical freeze-out. As the collision energy $\sqrt{s}$ increases, the chemical freeze-out temperature $T$ raises, while the baryon chemical potential $\mu_B$ decreases. Their behavior may be parameterized as follows~\cite{Cleymans:2005xv}:
\beqn
T(\mu_B) = a - b \mu_B^2 - c \mu_B^4,
\qquad
\mu_B(\sqrt{s}) = \frac{d}{1+f \sqrt{s}}, \qquad
\label{eq:parametrisation}
\eeqn
where the parameters $a,b,\dots,f$, determined from earlier results on the relativistic heavy-ion collisions, are given in Table~\ref{tbl:freezeout}.
\begin{table}[htb]
\begin{tabular}{c|c|c|c|c}
$a$, GeV & $b$, GeV & $c$, GeV & $d$, GeV & $f$, GeV${}^{-1}$ \\
\hline
$0.166(2)$ & $0.139(16)$ & $0.053(21)$ & $1.308(28)$ & $0.273(8)$
\end{tabular}
\caption{The parameters of the chemical freeze-out~\eq{eq:parametrisation}, from Ref.~\cite{Cleymans:2005xv}. The numbers in brackets give the error estimates in the last digit(s) of the parameters.}
\label{tbl:freezeout}
\end{table}

The predicted ratio ${\mathcal R}_{{\bar\Lambda}/\Lambda}$ of the spin polarizations~\eq{eq:R:definition} coming from the chiral-helical vortical effects~\eq{eq:R:prediction} is shown in Fig.~\ref{fig:R} along with the experimental data of the STAR collaboration~\cite{STAR:2017ckg}. The figure demonstrates a very good agreement between the analytical prediction and the experimental results, especially given the fact that no free fitting parameters were used to adjust the analytical curve.

At high collision energy, the chemical potential is much smaller compared to the temperature at the chemical freeze-out. Therefore, the asymmetry in the spin polarizations~\eq{eq:R:prediction} gets the following asymptotic form,
\beqn
{\mathcal R}_{{\bar\Lambda}/\Lambda} = 1 + \frac{8 \ln 2}{\pi^2} \frac{\mu_B}{T} + O(\mu_B^2/T^2).
\eeqn
and goes to unity (indicating the absence of asymmetry) in agreement with the experimental data on global polarization of hyperons. At the largest available data point for the STAR experiment~\cite{STAR:2017ckg}, $\sqrt{s_{NN}}=200\GeV$, the asymmetry in ${\bar\Lambda}/\Lambda$ polarizations is predicted, according to Eq.~\eq{eq:R:prediction}, to be less than $10\%$ in agreement with the experimental data which overlap within the error bars. At the ALICE energies, $2.76\,\mathrm{TeV}$ and $5.02\,\mathrm{TeV}$, the polarizations are close to zero~\cite{Acharya:2019ryw} and the asymmetry cannot be estimated reliably.

Evidently, our analysis is not restricted to the global properties of the rotating fluid since these considerations may also be repeated in local terms. In particular, it was recently observed by the STAR collaboration that the local vorticity of the Au-Au collisions at $\sqrt{s_{NN}}=200\GeV$ has a quadrupole component along the beam direction~\cite{Adam:2019srw}. The fits of the quadrupole signals of the spin-polarizations for hyperons and anti-hyperons by the sine function show that the magnitudes of these polarizations overlap with each other within $20\%$ errors. This result agrees with the range of our estimation~\eq{eq:R:prediction} as well.

%\vskip 1mm
\section{Electric current in noncentral collisions: chiral vs helical effects} 
\label{sec:electric}

\subsection{Helical vortical effect vs chiral vortical effect}

A vortical fermion fluid, in a state close to the thermal equilibrium, develops the vector (electric) current along the vorticity axis:
\beqn
{\bs J}_V = \frac{1}{\pi^2} \mu_V \mu_A {\bs \omega} + \frac{2 \ln 2}{\pi^2} \mu_H T {\bs \omega},
\label{eq:J:V}
\eeqn
where the first term corresponds to the well-known chiral vortical contribution~\cite{Vilenkin} while the second term gives the helical vortical effect~\cite{Ambrus:2019ayb,Ambrus:2019khr}. The axial ($\mu_A$) and helical ($\mu_H$) chemical potentials correspond to the fluctuations of the density of the axial charge and the density of the quark's helicity, respectively: $\rho_A = \mu_A T^2/3$ and  $\rho_H = \mu_H T^2/3$ (we neglected all sub-leading corrections to these formulas). In the state of thermal equilibrium, we expect that these densities vanish, $\mu_A = \mu_H = 0$.

The axial density measures the difference between left- and right-handed chiralities, while the helical density gives the difference between the contributions of quarks and anti-quarks to the axial density. These quantities share the obvious similarity (which often leads to their erroneous identification) and therefore the statistical fluctuations of their densities in the off-equilibrium plasma are expected to be of the same magnitude, $\mu_A \sim \mu_H$.

Let us compare the magnitudes of the vector current generated by the chiral and helical vortical effects~\eq{eq:J:V}. The difference may be quantified by the ratio of the corresponding prefactors in Eq.~\eq{eq:J:V} evaluated at the chemical freeze-out. For example, at the collision energy $\sqrt{s_{{NN}}}$ = 200 GeV, the fluctuations in the helical charge (in $\mu_H$) give a much larger contribution to the vector vortical conductivity compared to the fluctuations of the axial charge (in $\mu_A$) of the same magnitude (we take $\mu_A = \mu_H$):
\beqn
\frac{(J_V)_{\mathrm{HVE}}}{(J_V)_{\mathrm{CVE}}} = 6 \ln 2 \frac{T}{\mu_B} \simeq 30,
\quad  (\sqrt{s_{{NN}}} = 200 \GeV).
\qquad
\label{eq:win:1}
\eeqn
Therefore at these energies, the chiral vortical effect can be neglected in favor of the larger helical vortical contribution to the generated vector (electric) current~\eq{eq:J:V}. It is not difficult to verify that at lower collision energies, $\sqrt{s_{NN}} \sim$ 10 GeV, the chiral and helical conductivities give contributions to the vector current~\eq{eq:J:V} of the same order.

\subsection{Helical vortical effect vs chiral magnetic effect}

In addition to the vorticity in the quarks' fluid, the noncentral heavy-ion collisions create also a strong magnetic field ${\bs B}$ because the colliding ions carry an electrical charge. The fluctuations of the chiral (axial) charge density in the quark-gluon plasma, created in the collision, generate the electric (vector) current of quarks along the axis of magnetic field:
\beqn
{\bs J}_V = \frac{\mu_A}{2 \pi^2} e {\bs B}.
\label{eq:CME}
\eeqn
The transport law~\eq{eq:CME}, known as the chiral magnetic effect (CME), leads to potentially observable experimental consequences~\cite{Kharzeev:2015znc}. As the strength of the magnetic field rises with the collision energy, we consider below the energy $\sqrt{s_{{NN}}}$ = 200 GeV which favors the CME~\eq{eq:CME} and corresponds to the high-energy side of the energy scan of the RHIC facility.

In addition to the CME, the electric current can be generated by the chiral and helical vortical effects in the same collision~\eq{eq:J:V}. Since the helical vortical effect overwhelms its chiral counterpart, it is interesting to compare the efficiency of the CME and the HVE at high energies.

Qualitatively, it seems logical that the helical vortical effect can generate a stronger electric current as compared to the current created by the chiral magnetic effect. This conclusion is supported by the expected long-lasting nature of the vorticity as compared to the short relaxation time of the magnetic field~\cite{McLerran:2013hla}.

Quantitatively, we get the dominance of the helical vortical effect over the chiral magnetic effect:
\beqn
\frac{(J_V)_{\mathrm{HVE}}}{(J_V)_{\mathrm{CME}}} = 4 \ln 2 \cdot \frac{T \Omega}{e B} \simeq 3,
\quad  (\sqrt{s_{{NN}}} = 200 \GeV),
\qquad
\label{eq:win:2}
\eeqn
where we set $\mu_H = \mu_A$ following our earlier arguments. We also took the following estimations of the physical characteristics of the QGP at $\sqrt{s_{{NN}}} = 200 \GeV$: the chemical freeze-out temperature $T = 166\MeV$~\cite{Cleymans:2005xv}, the angular frequency $\Omega = 6.6\MeV$~\cite{STAR:2017ckg}, and the (optimistic) estimate of the magnetic field $B = 0.05 \, m_\pi^2$~\cite{McLerran:2013hla} of the electrically conducting QGP at the chemical freeze-out.

%\paragraph{Summary.} 

\section{Conclusions}

We have shown that the interplay of axial and helical vortical effects allows us to compute, separately, the spin polarizations of light quarks and light anti-quarks along the global vorticity axis in noncentral collisions. Assuming the spin-vector dominance in the transfer of the spin of light quarks to the spin polarization of the heavier strange quarks, we were able to express the ratio of the spin polarizations of $\bar\Lambda$ and $\Lambda$ via the ratio~\eq{eq:R:prediction} of the chiral~\eq{eq:sigmaA} and helical~\eq{eq:sigmaH} vortical conductivities evaluated at the chemical freeze-out. We show in Fig.~\ref{fig:R} that this much-simplified picture, that requires no fitting parameters, agrees very well with the experimental data of the STAR collaboration. 

We also noticed that the helical vortical effect should generate the electric current along the vorticity axis, which is comparable to (at low collision energies) and much larger than (at high energies) the current generated by the chiral vortical effect, assuming the same-order fluctuations of the helical and chiral densities. The helical vortical effect dominates also over the chiral magnetic effect at high energies in the RHIC range.

\acknowledgments
V.E.A. gratefully acknowledges the support of the Alexander von Humboldt Foundation through a Research Fellowship for postdoctoral researchers. The work of M.N.C. was partially supported by Grant No. 0657-2020-0015 of the Ministry of Science and Higher Education of Russia.

\end{document}